\documentclass[
    ,final            
  ]
  {aipproc}

\layoutstyle{6x9}

\begin{document}

\title{The Peculiar Type Ia Supernova 2005hk}


\author{V. Stanishev}{
  address={Physics Department, Stockholm University, AlbaNova University Center, 106 91 Stockholm, Sweden}
}

\author{S. Taubenberger}{
  address={Max-Planck-Institut f\"ur Astrophysik, PO Box 1317, 85741 Garching, Germany}
}

\author{G. Blanc}{
  address={Laboratoire Astroparticules et Cosmologie,   Universite Paris 7,
2 place Jussieu, 75005 Paris, France}
}

\author{G.C. Anupama}{
  address={Indian Institute of Astrophysics, Koramangala, Bangalore 560 034, India}
}


\author{S. Benetti}{
  address={INAF, Osservatorio Astronomico di Padova, vicolo dell'Osservatorio 5, 35122 Padova, Italy}
}

\author{E. Cappellaro}{
  address={INAF, Osservatorio Astronomico di Padova, vicolo dell'Osservatorio 5, 35122 Padova, Italy}
}

\author{N. Elias-Rosa}{
  address={INAF, Osservatorio Astronomico di Padova, vicolo dell'Osservatorio 5, 35122 Padova, Italy}
}

\author{C. F\'eron}{
  address={DARK Cosmology Center, Niels Bohr Institute, Copenhagen University Juliane Maries Vej 30, DK-2100 Copenhagen, Denmark}
}

\author{A. Goobar}{
  address={Physics Department, Stockholm University, AlbaNova University Center, 106 91 Stockholm, Sweden}
}

\author{K. Krisciunas}{
  address={Department of Physics, University of Notre Dame, 225 Nieuwland Science Hall, Notre Dame, IN
46556-5670}
}

\author{A. Pastorello}{
  address={Max-Planck-Institut f\"ur Astrophysik, PO Box 1317, 85741 Garching, Germany}
  ,altaddress={Department of Physics \& Astronomy, Queen's University,
  Belfast, BT7 1NN, Northern Ireland, UK} 
}
\author{D.K. Sahu}{
  address={Indian Institute of Astrophysics, Koramangala, Bangalore 560 034, India}
}

\author{M. E. Salvo}{
  address={Research School of Astronomy and Astrophysics, 
Australian National University, Mt Stromlo Observatory, 
via Cotter Road, Weston Creek PO 2611, Australia}
}

\author{B. P. Schmidt}{
  address={Research School of Astronomy and Astrophysics, 
Australian National University, Mt Stromlo Observatory, 
via Cotter Road, Weston Creek PO 2611, Australia}
}

\author{J. Sollerman}{
  address={DARK Cosmology Center, Niels Bohr Institute, Copenhagen University Juliane Maries Vej 30, DK-2100 Copenhagen, Denmark}
  ,altaddress={Department of Astronomy, Stockholm University, AlbaNova University Center, 106 91 Stockholm, Sweden} 
}

\author{C. C. Th\"one}{
  address={DARK Cosmology Center, Niels Bohr Institute, Copenhagen University Juliane Maries Vej 30, DK-2100 Copenhagen, Denmark}
}

\author{M. Turatto}{
  address={INAF, Osservatorio Astronomico di Padova, vicolo dell'Osservatorio 5, 35122 Padova, Italy}
}

\author{W. Hillebrandt}{
  address={Max-Planck-Institut f\"ur Astrophysik, PO Box 1317, 85741 Garching, Germany}
}

\begin{abstract}
We present a preliminary analysis of an extensive set of optical 
observations of the Type Ia SN\,2005hk. We show that the evolution of SN\,2005hk 
closely follows that of the peculiar SN\,2002cx. SN\,2005hk is more
luminous than SN\,2002cx, while still under-luminous compared to normal Type Ia supernovae. 
The spectrum at 9 days before maximum is dominated by conspicuous Fe~III and Ni~III lines, 
and the Si~II\,6355 line is also clearly visible. All these features have 
low velocity ($\sim6000$\,km/s). The near maximum spectra show lines of 
Si~II, S~II, Fe~II-III, as well as other intermediate mass and iron group elements.
Analysis with the code for synthetic spectra SYNOW indicates that all these spectral 
lines have similar velocities.
 \end{abstract}

\maketitle

\section{Introduction}

Type Ia Supernovae (SNe Ia) form a fairly homogeneous class of objects
\citep[see, e.g.][]{branch_snei_rev}. Most of them fall into the so-called "Branch-normal" 
group \citep{branch_pec}. However, spectroscopically peculiar SNe have also 
been found, 
and their fraction might be non-negligible \citep{branch_pec,li_pec}. The peculiar SNe\,Ia
are classically divided into two groups -- 1991bg-like and 1991T-like \citep[see][]{li_pec}.
Recently, some objects that do not fit in any of these categories  
have been discovered \citep[see, e.g.][]{li_02cx} and 
included into a group with  SN\,2002cx as prototype \citep{jha_02cx}. 

SN\,2005hk was discovered on Oct. 30.25 UT in UGC\,272 by \citet{05hk_disc}. 
We obtained two spectra
of SN\,2005hk on  Oct. 31st. The spectral features were similar to
those observed in the 1991T-like events, about 10 days before maximum light.
After this, an observational campaign
was started by the European Supernova Collaboration (ESC) 
and an extensive set of 
data was obtained using a number of telescopes.
A more thorough study of the early spectra, however, revealed that SN\,2005hk 
was more similar to SN\,2002cx. Thus, 
SN\,2005hk became the best observed 2002cx-like event so far. In this 
work we present a preliminary analysis of a significant part of our data.
 
\section{Results}

\begin{figure}[!t]
  \includegraphics*[width=0.95\textwidth]{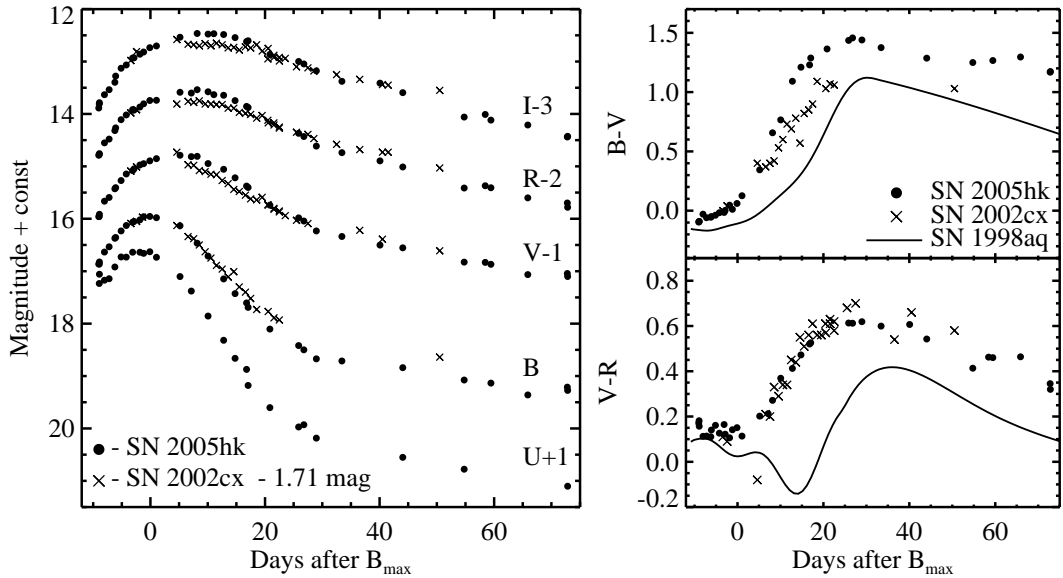}
  \caption{The $UBVRI$ light curves of SN\,2005hk (left), 
  and the $B-V$ and $V-R$ color indexes of SN\,2005hk (right) compared to those of 
SN\,2002cx and the normal  SN Ia 1998aq \citep{riess_98aq}.}
  \label{f:phot}
\end{figure}

After the standard bias and flat field corrections, the magnitudes of
SN\,2005hk were measured with the point-spread function (PSF) fitting 
technique. The instrumental magnitudes were transformed to Johnson-Cousins standard
magnitudes using field stars whose magnitudes were calibrated 
on two photometric nights at CTIO. The 
photometry was calibrated using linear equations derived from observations
of \citet{landolt92} stars. The $UBVRI$ light curves, and the $B-V$ and
$V-R$ color indexes are shown in Fig.\,\ref{f:phot}. The light curves show well-defined 
single maxima in all bands.
SN\,2005hk reaches the $B$-band maximum light $B_{\rm max}=15.96$ mag on Nov. 9th
(JD\,2453684.5$ \pm0.5$) and $\Delta M_{15}(B)\simeq1.5$  was estimated. 
The $UVRI$ peak magnitudes of 15.62, 15.80, 15.57 and 15.46 mag 
were reached $-1.6, +2.2, +7.5$ and $+10$ days from the $B$ maximum,
i.e.  the light curve maxima occur progressively later from $U$ to $I$-band. 
The peak magnitudes are
uncertain to $\sim0.02$ mag and the epochs of the maxima to $\sim0.5$ days.
The light curves of SN\,2002cx \citep{li_02cx} are also shown in 
Fig.\,\ref{f:phot}, shifted by 1.71 mag to match the $B$ band peak
magnitude of SN\,2005hk. The light curves of 
the two SNe are very similar; however,  a few differences are noticeable. 
The $R$ and $I$ band maxima of SN\,2002cx are flatter than those of 
SN\,2005hk. The color indexes of 
the two SNe are similar, and generally much redder than those of normal SNe Ia.

\begin{figure}
  \includegraphics*[width=0.95\textwidth]{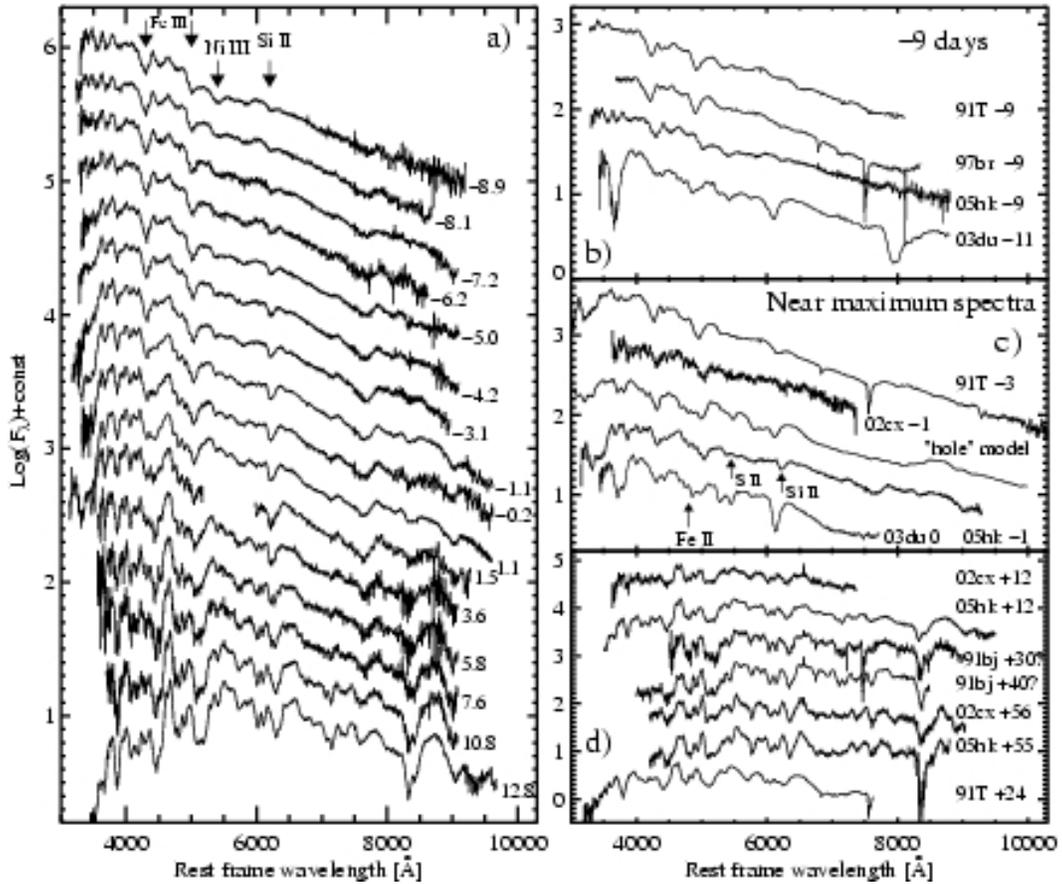}
  \caption{{\bf (a)} Early-time spectral evolution of  SN\,2005hk and {\bf (b-d)} comparison with
spectra of  SN\,1991T  \citep{pilar_91t,phillips_91t}, 1997br \citep{li_97br}, 
 2003du \citep{me_03du}, 1991bj \citep{gomez_91bj,capp_91bj} at different epochs. 
A "hole" model \citep{kasen_hole} (see text) is also shown
in (c) panel.}
  \label{f:sp}
\end{figure}

The early-time spectral evolution of  SN\,2005hk is shown in Fig.\,\ref{f:sp}a. 
The -9 days spectrum is dominated by 
Fe~III and Ni~III lines, and only weak Ca II H\&K and Si II 6355 lines are visible.
The spectrum is very similar to the 91T-like SNe Ia (Fig.\,\ref{f:sp}b),
suggesting a rather high temperature, while the line velocities are
anomalously low, only $\sim6000$ km/s. With time, lines of intermediate mass elements 
start to emerge and by the time of maximum Si II, S II, Mg II and Ca II 
are clearly detected together with the Fe~III lines. Fe~II lines around 4800-5200\,\AA\ are also 
prominent (Fig.\,\ref{f:sp}c). 
In general, around maximum the spectrum of SN\,2005hk 
is not much different from that of a normal SN Ia (e.g. SN\,2003du), but the lines of
intermediate mass elements are  considerably weaker. After maximum, the spectral evolution of SN\,2005hk  
closely follows SN\,2002cx (Fig.\,\ref{f:sp}d), which is different from all
other SNe Ia sub-types. 
According to \citet{branch_02cx}, the post-maximum spectra 
of SN\,2002cx are dominated mostly by Fe~II absorption lines.
In addition, in Fig.\,\ref{f:sp}d we show two spectra of  SN\,1991bj 
\citep{gomez_91bj,capp_91bj}, which we have identified as a 2002cx-like discovered approximately a month after maximum.

\begin{figure}
  \includegraphics*[width=\textwidth]{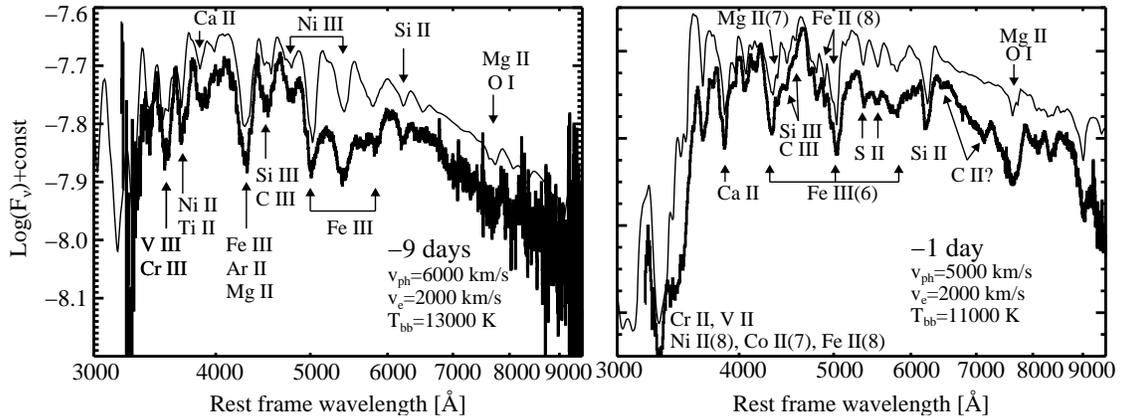}
  \caption{The $-9$ and $-1$ days spectra of SN\,2005hk compared 
with SYNOW synthetic spectra (for clarity, the model spectra 
are shifted up by 0.05 dex). The SYNOW global parameters and the 
species contributing to the absorption features are indicated. The numbers 
reported in brackets after the name of the ions is 
the velocity (in units of 1000\,km/s) at which the corresponding ion is detached.}
\label{f:synow}
\end{figure}

We used the supernova spectral synthesis code SYNOW 
\citep[for details see, e.g.][]{branch_98aq} to give more 
insights to the properties of SN\,2005hk spectra. We modeled spectra at two epochs, 
$9$ and $1$ days before the $B$ maximum. The synthetic spectra (for clarity shifted up by 0.05 dex) and 
the adopted parameters are shown in Fig.\,\ref{f:synow}. 
The adopted excitation temperature  was 10000\,K for the singly and 15000\,K for the doubly
ionized species. Note that besides the great simplifications assumed by SYNOW, some
of the derived parameters are correlated, e.g. the photospheric $v_{ph}$ and the e-folding
velocity $v_{e}$. 

The species contributing to the 
strongest features in the spectra are labeled in Fig.\,\ref{f:synow}. In the  $-9$ days spectrum
we consider the presence of Fe~III, Ni~III and Si~II lines as definite, 
and Ca~II as likely. The feature 
around 4500\AA\ is best reproduced with Si~III and C~III. The features in the 
blue part of the spectrum are due to singly and doubly ionized Fe-group elements.
Most notably, the strong feature around 3600\AA\ is better fitted with V~III and Cr~III. 
It was also found that Ar~II improves the fit of the strong 4300\AA\ feature; the absorption 
is too shallow if only Fe~III is used. Although the synthetic spectrum reproduces fairly well
most of the observed features, there are wavelength intervals which are not well fitted. 
For example, the absorption near 3900\AA\ is too broad to be attributed to Ca~II H$\&$K only.
The observed flux level between 5000\AA\ and 6000\AA\ cannot be reproduced and the 
synthetic spectrum gives too much flux.

Most of the observed absorption features in the $-1$ day spectrum are well 
reproduced by the synthetic spectrum. The photospheric velocity used is 
5000\,km/s, but it was found that many
of the lines should be detached. In addition to the species used to model the
$-9$ days spectrum, for the  $-1$ day synthetic spectrum S~II, Fe~II, Cr~II, V~II and Co~II were
added, and no Ni~III, Cr~III or V~III was used. The line around 7200\AA\ could be 
due to C~II; however, as discussed by \citet{chor_05hk} if this was the case, 
there should be a stronger feature around 6400\AA\ which is not visible in the 
observed spectrum. 
It should also be noted that Ar and V are typically not considered
in the analyzes of SN\,Ia spectra and their presence in SN\,2005hk is not certain.

\section{Discussion}

The recession velocity of UGC~272 corrected for 
the Local Group in-fall onto Virgo is 3863\,km/s 
(from the LEDA database\footnote{\url{http://leda.univ-lyon1.fr}}), providing
a distance modulus of $\mu=33.65$ mag (assuming $H_0=72$\,km/s/Mpc).
The observed $B_{\rm max}=15.96$ mag of SN\,2005hk thus implies an 
absolute magnitude $M_B\simeq-17.7$. 
The Milky Way reddening in the direction to SN\,2005hk is only $E(B-V)=0.023$ \citep{mw_red}
and the host galaxy reddening is $E(B-V)=0.091$ \citep{chor_05hk}. 
The total dimming by dust in the $B$ band is thus 
0.47 mag (with $R_B=4.1$). SN\,2005hk is therefore
significantly sub-luminous compared to the 
normal SNe\,Ia with $\Delta M_{15}(B)\sim1.5$, whose expected absolute magnitude is
$M_B\simeq-19$ \citep[e.g.,][]{prieto}.
\citet{li_02cx} estimated that SN\,2002cx had $M_B\simeq-17.55$ and 
$\Delta M_{15}(B)\simeq1.3$\footnote{$\Delta M_{15}(B)$ of SN\,2002cx might actually be 
$\simeq1.6$ according to \citet{li_02cx}}. With $M_B\simeq-18.2$, SN\,2005hk is thus 
more luminous than SN\,2002cx.

The  $UBVRI$ magnitudes were transformed to fluxes using the absolute calibration of \citet{bessell98}
and then 
integrated 
in order to estimate the $uvoir$ bolometric luminosity of SN\,2005hk. A correction 
for the flux emitted in the near-infrared part of the spectrum was also applied. 
This correction, $\sim10$\% at 
maximum and  $\sim42$\% one month after, was derived integrating the combined 
optical and near-infrared spectra at five epochs during the first month after $B$ band maximum.
The $uvoir$ bolometric maximum of 4.24$\times10^{42}$\,erg/s occurred $\sim3$ days after the 
$B$-band maximum.

With its low luminosity and peculiar light-curve morphology with single maxima occurring
progressively later in the red bands,  SN\,2005hk is more similar to Type Ic SNe
rather than SNe\,Ia. The time evolution of the color indexes of 
SN\,2005hk is also quite similar to SNe\,Ic \citep[see e.g.,][]{taub_04aw}.
 While the broadband colors evolved similarly to SNe\,Ic, the spectral features' 
evolution was closer to SNe\,Ia and suggests that SN\,2005hk was likely a 
thermonuclear event. In favor of this is the clear detection
of Si and S lines in the near maximum spectra.
We have also obtained a late-time spectrum of SN\,2005hk, which is quite similar 
to the spectrum of SN\,2002cx at a similar epoch \citep{jha_02cx}. 
According to \citep{jha_02cx} the late-time spectrum
is dominated by Fe absorption lines \citep{jha_02cx}, which also suggests 
a thermonuclear explosion. 
In contrast, 
late-time spectra of SNe\,Ic are dominated by forbidden emission lines of O, Ca and Mg.

The studies of SNe\,2002cx and 2005hk \citep{li_02cx,branch_02cx,jha_02cx,chor_05hk}
point out that no existing SN model accounts for the peculiar properties of the 
2002cx-like SNe. In particular, it is difficult to explain the combination of hot early-time spectra, 
and low luminosity and expansion velocities. The observed polarization 
of SN\,2005hk is relatively low \citep{chor_05hk} and models involving large asymmetry 
are unlikely. The explosion simulations of \citet{marietta} show that the interaction 
 with a companion star may leave a hole of the ejecta.
\citet{kasen_hole} computed flux and polarization spectra, and the luminosities of such
a model with different orientations of the hole.
They suggested that SN\,2002cx might be a 1991bg-like, but seen straight down the hole. 
This allows to see into the deep hot layers and  the synthetic
spectra  resemble the early-time spectra of the 1991T-like SNe, but have 
low expansion velocity and luminosity. 
Moreover, when looking  straight down the hole the polarization will be small. 
In Fig.\,\ref{f:sp}c we compare 
the near-maximum spectrum of SN\,2005hk with a synthetic spectrum from \citep{kasen_hole} 
when looking directly into the ejecta hole. Although, the two spectra are 
similar, recent studies cast doubts on this scenario \citep{jha_02cx,chor_05hk}.

The late-time spectra of 2002cx-like SNe are completely different from all other
supernova types, which let \citep{jha_02cx} to suggest that radically different models
may be needed. 
Our Fig.\,\ref{f:sp}d and Fig.\,5 in \citet{jha_02cx}
clearly demonstrate the spectral homogeneity within this group and strengthen 
the hypothesis that these objects belong to a different SN type.

\begin{theacknowledgments} 
This work is supported in part by the European Community's Human
Potential Program ``The Physics of Type Ia Supernovae'', under
contract HPRN-CT-2002-00303.  V.S. and A.G. would also like to thank
the G\"oran Gustafsson Foundation for financial support. We are grateful 
to the numerous observers who gave up part of their time to 
observe SN\,2005hk and the support astronomers of the telescopes for performing
part of the observations. We thank Jan-Erik Ovaldsen for reducing part of the 
photometric data, Dan Kasen for making available to us his "hole " model
synthetic spectra and David Branch for allowing us to use SYNOW.
\end{theacknowledgments}

\bibliographystyle{aipproc}   

\end{document}